\documentclass[
]{ceurart}

\usepackage{listings}
\usepackage{adjustbox}
\usepackage[most]{tcolorbox}
\begin{document}

\copyrightyear{2025}
\copyrightclause{Copyright for this paper by its authors.
  Use permitted under Creative Commons License Attribution 4.0
  International (CC BY 4.0).}

\conference{CLEF 2025 Working Notes, 9 -- 12 September 2025, Madrid, Spain}

\title{Beyond Retrieval: Ensembling Cross-Encoders and GPT Rerankers with LLMs for Biomedical QA}

\title[mode=sub]{Notebook for the BioASQ Lab at CLEF 2025}


\author[1]{Shashank Verma}[
    orcid=0009-0003-1478-4328,
    email=sverma342@gatech.edu,
]
\cormark[1]
\fnmark[1]

\author[1]{Fengyi Jiang}[
    orcid=0009-0007-9420-5259,
    email=art.jiang@gatech.edu,
]
\fnmark[1]

\author[1]{Xiangning Xue}[
    orcid=0000-0002-5775-1193,
    email=xxue47@gatech.edu,
]
\fnmark[1]

\address[1]{Georgia Institute of Technology, North Ave NW, Atlanta, GA 30332}
\cortext[1]{Corresponding author.}
\fntext[1]{These authors contributed equally.}

\begin{abstract}
  Biomedical semantic question answering rooted in information retrieval can play a crucial role in keeping up to date with vast, rapidly evolving and ever-growing biomedical literature. A robust system can help researchers, healthcare professionals and even layman users access relevant knowledge grounded in evidence. The BioASQ 2025 Task13b Challenge serves as an important benchmark, offering a competitive platform for advancement of this space.
    This paper presents the methodologies and results from our participation in this challenge where we built a Retrieval-Augmented Generation (RAG) system that can answer biomedical questions by retrieving relevant PubMed documents and snippets to generate answers.
    For the retrieval task, we generated dense embeddings from biomedical articles for initial retrieval, and applied an ensemble of finetuned cross-encoders and large language models (LLMs) for re-ranking to identify top relevant documents. Our solution achieved an MAP@10 of 0.1581, placing 10th on the leaderboard for the retrieval task.
    For answer generation, we employed few-shot prompting of instruction-tuned LLMs. Our system achieved macro-F1 score of 0.95 for yes/no questions (rank 12), Mean Reciprocal Rank (MRR) of 0.64 for factoid questions (rank 1), mean-F1 score of 0.63 for list questions (rank 5), and ROUGE-SU4 F1 score of 0.29 for ideal answers (rank 11).
\end{abstract}


\maketitle

\section{Introduction}
Answering biomedical questions with precise, well supported responses requires intensive searching, reading and synthesizing information from dozens of research articles - a time consuming task demanding both domain expertise and contextual understanding. It poses unique challenges due to domain-specific jargon and compositional question structure. Solving this problem will enable access to an accurate biomedical assistant for researchers and clinicians, supporting their decision-making and work efficiency and could even open doors for access to general public for self education and to help fight biomedical misinformation. BioASQ Task 13b Biomedical Semantic QA \cite{BioASQ2025task13bSynergy} describes this exact challenge, dividing it into two phases: Phase A is a retrieval task that identifies the most relevant list of documents and snippets for the given question, while Phase B focuses on generating both a formatted \textit{exact answer} and a natural language \textit{ideal answer} using the documents \& snippets as context.


\begin{figure}[t]
    \centering
    \includegraphics[width=0.4\linewidth]{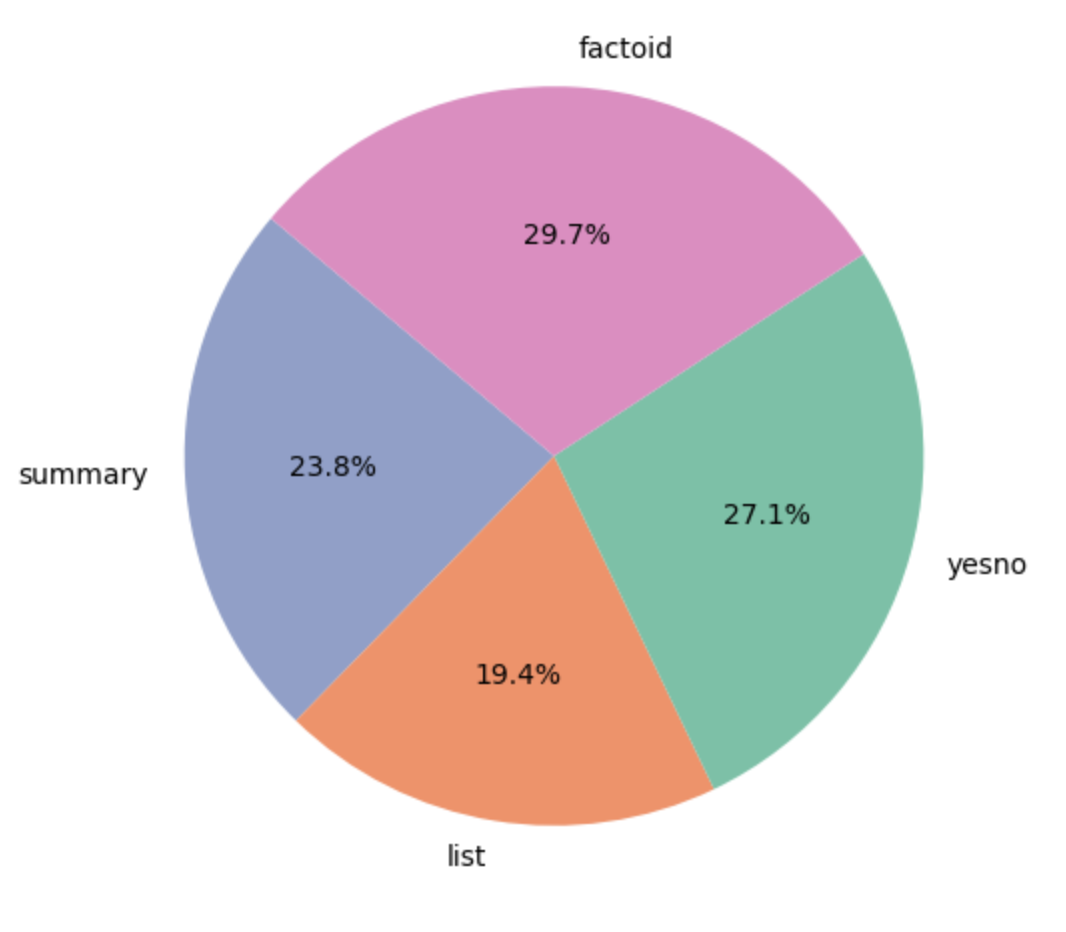}
    \includegraphics[width=0.55\linewidth]{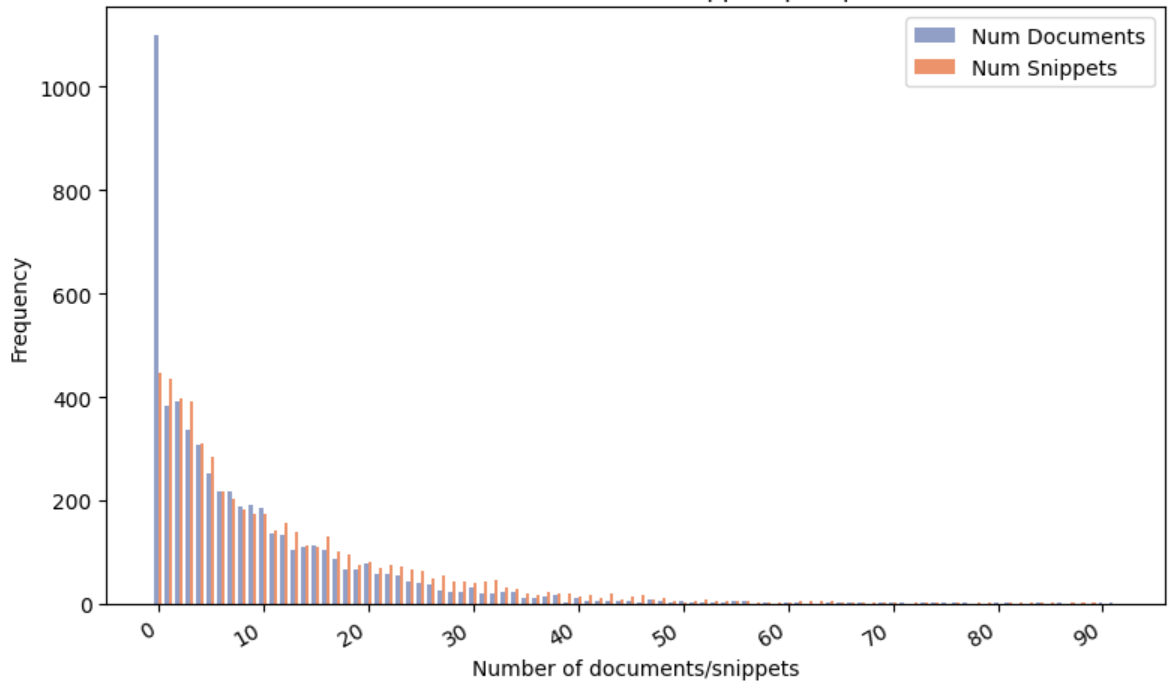}
    \caption{Distribution of question types in BioASQ training data (left) and number of golden documents \& snippets per question (right).}
    \label{fig:question-type-dist}
\end{figure}

The BioASQ challenge (Nentidis et al.~\cite{BioASQ2025overview}) has been continuously evolving to advance biomedical question answering, and we are participating in its 13th edition. We were provided a BioASQ-QA dataset \cite{QAcorpusBioASQ} with 5,389 questions, each accompanied with gold responses for most relevant documents, snippets, exact answers, and ideal answers that were curated by field experts. The number of gold documents and snippets varied but were generally in the order of tens (Fig. \ref{fig:question-type-dist}). Questions were divided into four types as shown in Fig. \ref{fig:question-type-dist}: yes/no (n=1,459), factoid (n=1,600), list (n=1,047), and summary (n=1,283). Question types define the format of exact answers: ``yes'' or ``no'' for yes/no questions; factoid questions expect a ranked list of alternative correct answers ordered by confidence, whereas list questions require a complete set of items without ranking. In addition to the exact answers, each of the question types also contain an ideal answer which is a natural language response to the question.

In this paper, we present our automated Retrieval-Augmented Generation (RAG) system, which aims to solve this problem. While prior systems have largely relied on sparse retrieval using BM25~\cite{robertson1995okapi}, which rely on word matching or entity based searching through PubMed API, our approach utilizes modern dense retrieval techniques and neural re-rankers with advanced LLM-based generation, allowing for strong performance evidenced by consistent top10 placement in the leaderboard across phases and question types. In summary, our work makes the following contributions:

\begin{enumerate}
    \item We built a dense retrieval system using \texttt{bge-large-en} embeddings on the full PubMed corpus, significantly improving document recall.
    \item We designed an ensemble of finetuned cross-encoders and GPT-based re-rankers to maximize MAP@10 for retrieved documents.
    \item We implemented a few-shot prompting pipeline with instruction-tuned LLMs to generate both exact and ideal answers.
\end{enumerate}


\section{Related Work}

\subsection{Retrieval-Augmented Generation (RAG)}
Retrieval-augmented generation (RAG) has emerged as a pivotal framework in natural language processing, particularly for enhancing the performance of large language models on knowledge-intensive tasks. The seminal work by Lewis et al \cite{lewis2020retrieval}, introduced this paradigm, demonstrating how combining retrieval-based and generation-based approaches can improve accuracy by grounding model outputs in external documents.
This approach is especially relevant for domains requiring up-to-date and domain-specific knowledge, such as biomedical question and answering. 
Subsequent work by Izacard et al \cite{izacard2021leveraging} further refined RAG by introducing Fusion-in-Decoder (FiD), which enhances generation by aggregating information from multiple retrieved documents.

\subsection{Vector Databases for Scalable Search}
Storing and querying large sets of embeddings, requires scalable and efficient data structures. 
Malkov et al\cite{malkov2016efficient} introduced Hierarchical Navigable Small World (HNSW) algorithm  and provided the foundation of many vector database implementations, offering logarithmic complexity scaling for similarity searches. Johnson et al\cite{johnson2017billion} further explored billion-scale similarity search, emphasizing GPU acceleration.
Additionally, Annoy by Bernhardsson \cite{bernhardsson2018annoy} offers an alternative ANN approach with tree-based indexing, though we prioritize HNSW for its balance of speed and accuracy.

\subsection{Retrieval with Bi-Encoders and Cross-Encoders}
Efficient and accurate retrieval is central to RAG pipeline. Common ranking methods include traditional statistical models such as BM25~\cite{robertson1995okapi}, which rely on word matching, and more parameterized deep learning models like transformers based cross-encoders and bi-encoders, examples including PubMedBERT~\cite{gu2021domain}, MiniLM~\cite{wang2020minilm}.
Bi-encoders, such as Sentence-BERT (Reimers and Gurevych \cite{reimers2019sentence}), encode queries and documents independently into dense vectors, enabling efficient similarity searches over large datasets.
However, bi-encoders often trade precision for speed.
Rosa et al\cite{rosa2022defense} demonstrated the superiority of cross-encoders in zero-shot retrieval tasks while Nogueira and Cho \cite{nogueira2019passage} explored their use in passage re-ranking with BERT.
Retrieving large number of documents with a bi-encoder and re-ranking to the top-k with a cross-encoder mirrors strategies evaluated in the BEIR benchmark by Thakur et al\cite{thakur2021beir}. 

\subsection{LLMs for Re-ranking}
Large Language Models (LLMs) are state of the art in many natural language processing tasks and have been employed in various aspects of Information Retrieval including passage re-ranking. Recent work by Ma et al.~\cite{ma2023zeroshotlistwisedocumentreranking} and Pradeep et al.~\cite{pradeep2023rankvicunazeroshotlistwisedocument} has shown that LLMs are able to perform zero-shot listwise document re-ranking leveraging their unparalleled understanding of natural language and reasoning capabilities, somewhat bypassing the need for domain-specific fine-tuning.


\subsection{Fine-Tuning Domain-Specific Language Models}

Luo et al~\cite{luo2022biogpt} introduced BioGPT with domain-specific pre-training on 15 million PubMed abstracts which makes it well-suited for biomedical text generation. It can be adapted efficiently using Low-Rank Adaptation (LoRA) by Hu et al~\cite{hu2021lora}, which reduces the number of trainable parameters by introducing low-rank updates to pre-trained weights. This parameter-efficient method mitigates the computational constraints we encountered, such as memory limitations, while maintaining performance. 

Complementary approaches like AdapterFusion by Pfeiffer et al\cite{pfeiffer2020adapterfusion} offer modular fine-tuning across tasks, though we favor LoRA for its simplicity and compatibility with BioGPT. Lee et al\cite{lee2020biobert} also highlight the value of domain-specific pre-training with BioBERT, reinforcing the importance of tailoring models to biomedical contexts.

\subsection{Few-Shot Learning with Large Language Models}

Few-shot prompting using instruction-tuned models like GPT-4o-turbo~\cite{achiam2023gpt} and Mistral-7B-Instruct-v0.3~\cite{chaplot2023albert} leverages in-context learning, a concept popularized by Brown et al~\cite{brown2020language} with GPT-3, where models generate outputs based on a few prompt examples.
Wei et al.~\cite{wei2022chain} extended this with chain-of-thought prompting, improving reasoning in complex tasks. 
Ateia and Kruschwitz~\cite{ateia2024can} developed frameworks for structured answer generation in specialized domains.

\section{Methodology}

\subsection{Phase A - Indexing \& Retrieval}
To address the limitations of PubMed API’s keyword-based retrieval and improve recall, we built a custom index powered by dense vector search.

\subsubsection{Dataset Preparation: PubMed 2025 Baseline}
We downloaded the full PubMed 2025 Baseline \cite{pubmedBaseline2025}, containing approximately 39 million records, including both article metadata and abstracts. The initial corpus was highly heterogeneous, with a subset of articles missing abstracts. Since abstracts are critical for semantic retrieval and question answering, we performed an initial preprocessing step to remove the documents with empty or missing abstracts, de-duplicated documents and filtered out the corrupted records.
We then parsed the XML to extract the PMID, title and abstract \cite{pubmedBaseline2025}.
After filtering, the resulting corpus contained $\approx$34 million high-quality abstracts.
\subsubsection{Embedding Generation: Bi-Encoder Model}
For dense semantic indexing, we employed a bi-encoder architecture, wherein documents and queries are encoded independently into vector representations, and similarity is computed via a simple metric such as cosine similarity.

We selected the \texttt{bge-large-en} model, a state-of-the-art bi-encoder trained for English retrieval tasks, due to its large model capacity (1024-dim embeddings) and open-source availability.
Each PubMed abstract was tokenized and encoded into a fixed-length 1024-dimensional embedding vector with care to truncate the abstracts to model's maximum token limit of 512, if necessary.

\subsubsection{Vector Databases for Semantic Search}

The full set of embeddings was stored in a vector database optimized for large-scale approximate nearest-neighbor (ANN) search. During retrieval, given a user query, the query was similarly embedded and the top-K most similar documents were retrieved based on cosine similarity in the embedding space.

Initially, we explored using fully managed cloud-based deployment like Pinecone, but ultimately discarded it due to its high cost \cite{pineconePricing2025}. We finally settled on using a local Qdrant deployment \cite{qdrant2025} which allowed us to gain full fine-grained control over our index while saving the hosting cost of Pinecone.

\begin{figure}[b]
    \centering
    \includegraphics[width=1\linewidth]{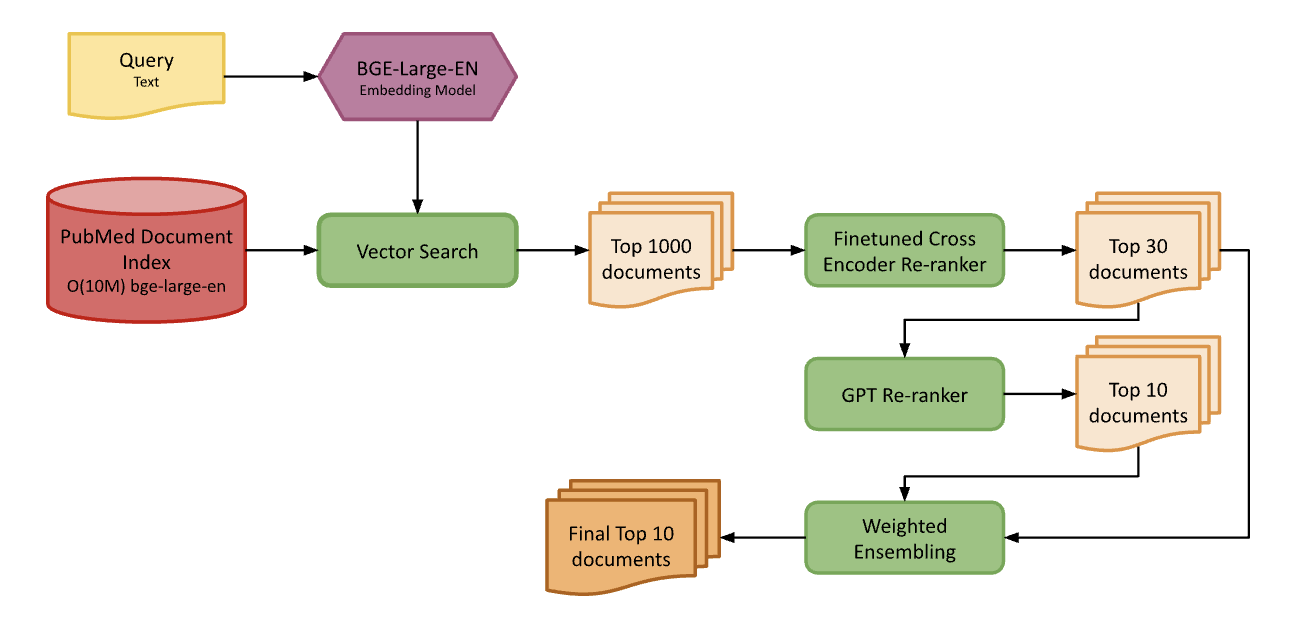}
    \caption{Architecture of the best performing system for Phase A. We use bge-large-en for creating our index and perform ensembled re-ranking to get the top-10 most relevant documents for a query.}
    \label{fig:phasea-architecture}
\end{figure}

\subsection{Phase A - Document \& Snippet Re-ranking}
In a typical Retrieval-augmented generation (RAG) system, the quality of the retrieved content as context directly influences the accuracy and relevance of the generated responses. With a high-quality index in place, we can retrieve the top 10 documents and snippets by directly querying the index using the question embedded with the same \texttt{bge-large-en} model. However, as we discussed before, retrieving a large pool of documents followed by a re-ranking of those documents leads to better results~\cite{10.1145/3269206.3271800}.

Due to time constraints, we weren't able to replicate our document retrieval \& re-ranking setup as demonstrated in Fig. \ref{fig:phasea-architecture} for snippets. Across all systems, we used the same snippet retrieval strategy (which was the predecessor of all future strategies).

\subsubsection{Cross Encoders for Re-ranking}
In our systems, we retrieve top-1k documents and then apply a re-ranking step to select the final top-10 documents. Bi-encoders embed question and document independently and compute similarity based on their embeddings which is the secret to their efficiency but also doesn't allow them to exploit the interaction between the question and the document. This is addressed by using cross-encoders which jointly process the question and candidate document pairs.


Due to slower speed of cross-encoders, the two-stage hybrid approach of bi-encoders to maximize recall with top-1k retrieval followed by cross-encoders to maximize MAP@10 in the final top-10 documents is the optimal approach balancing accuracy, performance, and computational cost.

\subsubsection{Finetuning Cross Encoders}
Every dataset and domain has certain nuances that are unique to it, which means that a general purpose model cannot exploit them. This is where a model finetuned on the domain's datasets can outperform an out-of-the-box model. Since we had the BioASQ-QA dataset provided by the competition, we finetuned a few cross encoder models to learn the intricacies of PubMed knowledge base and how the field experts perceive them.

We stratified the dataset on question to split for train, val and test, to make sure that the questions evaluated on the validation and test set were never seen by the model during training. Positive examples were created using the golden documents and negative examples were generated by querying the index and removing the golden documents. These served as hard negatives for the model since they were somewhat relevant to the question, allowing the model to learn the fine line between somewhat relevant and truly relevant. Training dataset was prepared as question-document pair as input and a label of 0/1 denoting whether the document was in the golden set for the question or not. The model was tasked with predicting the label given a question-document pair using a binary cross entropy loss. Evaluation and test data were prepared as tuples of \textit{question}, \textit{all\_documents} and \textit{golden\_documents}, letting the model apply re-ranking on \textit{all\_documents} to select the top-10 and be evaluated on metrics like MAP (mean average precision) which is what our competition was going to use on the leaderboard. We didn't keep a minimum relevance threshold for the re-ranker output, and always selected the top-10 most relevant candidates.

\noindent Finally, we experimented with some dataset variations:
\begin{itemize}
    \item \textbf{Equal ratio of positive and negative examples}: Ultimately discarded, since in the real application, we want to apply re-ranking to 1k documents to get top-10 which means positive and negative examples shouldn't be equal.
    \item \textbf{Discarding older questions from training}: This was done with the rationale that there is an inevitable data drift with any temporally collected data. Models trained on recent data could better inform the future predictions. This turned out to be true in our case and we used the model trained only on last few years' data.
\end{itemize}

\subsubsection{LLMs for Re-ranking}
LLMs understand the nuances of a text better than any language models. With that in mind, we experimented with using LLMs like gpt-4o for the re-ranking task. We explored two approaches:

\begin{itemize}
    \item  Prompt with the question-candidate pair and ask it to assign a relevance score. Then return the top 10 most relevant documents.

    \item  Prompt with the question and all the top-30 candidate documents and ask it to return the re-ordered list of top-10 documents. In this scenario, we used the top-30 from applying the finetuned cross-encoder on the retrieved top-1k.
\end{itemize}

The first approach resulted in lower MAP@10 compared to finetuned cross-encoder re-ranker. However, the second approach led to a sizable improvement.

\subsubsection{Ensembling}
We had two re-ranking systems: a) System1 was using finetuned cross encoder to get top-30 documents from 1k retrieved candidates, and b) System2 used finetuned cross-encoder to get to top-30 and gpt-4o prompting to further refine to top-10. We explored a couple of ways of combining the two systems:
\begin{itemize}
    \item \textbf{Nomination-based approach}: First 6 from \textit{System1}  and remaining 4 from \textit{System2}.
    \item \textbf{Weighted combination of scores}: A weighted summation of both system's documents. Grid search was employed to find the best weights.
\end{itemize}

Weighted Ensembling gave us the best MAP of all the systems. The final architecture of our best solution using weighted ensembling for Phase A is shown in Fig. \ref{fig:phasea-architecture}.

\subsection{Phase B System: Answering Generation with Retrieved Data}

In Phase B, we focused on generating answers using few-shot prompting of instruction-tuned large language models (LLMs), leveraging snippets retrieved in Phase A. Specifically, GPT-4o-turbo and Mistral-7B-Instruct-v0.3 models were used, implementing a variety of prompting templates designed to effectively guide the answer generation. While we also conducted experiments with fine-tuning BioGPT with task-specific heads, these yielded less competitive results compared to our prompting strategy; thus, detailed fine-tuning outcomes are omitted here but will be publicly available on our GitHub repository to support reproducibility and transparency.

\begin{figure}[htbp]
\centering
\begin{tcolorbox}[width=0.9\linewidth, colback=gray!5, colframe=black]
\large 
\textbf{Prompt Template}\\
\small
System: You are a biomedical assistant. You answer questions based on background knowledge and return the answer in the requested format.\\
\textbf{(Few-shot examples, repeated $\{n\}$ times)}\\
User: \{Query Template\}\\
Assistant: \{Answer of the query\}\\
\textbf{(Final query)}\\
User: \{Query Template\}\\
\end{tcolorbox}
\caption{Overall Prompt Template Structure. Template texts wrapped in \{\} are subject to change according to styles and questions. }
\label{fig:prompt-structure}
\end{figure}

\begin{figure}[htbp]
\centering
\begin{tcolorbox}[width=0.9\linewidth, colback=gray!5, colframe=black]
\large 
\textbf{Query Template 1}\\
\small
Passage: \{All snippets\} \\
Question: \{Question body\} \\
\{Answer Formatting Template, e.g. Answer this question in ``yes'' or ``no''. \} \\

\large 
\textbf{Query Template 2} (changed from Template 1 only for ideal answer prompts:)\\
\small
Passage: \{All snippets\} \\
Question: \{Question body\} (Hint: short answer is \{System generated exact answer\} ) \\
\{Answer Formatting Template\} 
\end{tcolorbox}
\caption{Query Templates for exact and ideal answers.}
\label{fig:query-templates}
\end{figure}

\begin{figure}[htbp]
\centering
\begin{tcolorbox}[width=0.9\linewidth, colback=gray!5, colframe=black]
\large 
\textbf{Answer Formatting Template 2} (showing key changes from Template 1)\\
\small
\textbf{List: } \\
- Answers should be based on the passage content, but you may supplement with your own knowledge if necessary. \\
- Prefer the shortest, cleanest, and most direct entry cut from the Passage that -accurately answer the question. \\
- Note that all phrases, even not complete, are related to the question. \\
\textbf{Factoid: } \\
- Each element must be a complete answer to the question. Do not include synonyms or duplicate meanings. Avoid acronyms unless the full name is unavailable. \\
Avoid overly long phrases, redundant details, or overly specific descriptions unless necessary. \\
- When multiple numerical answers are given (e.g., different rates, frequencies, or ranges across populations), merge them into a single range or compact expression if appropriate, instead of listing them separately. \\
- The returned entity names must not be synonyms of each other. \\
\textbf{Summary: } \\
- Prefer answers of only combining direct phrases or sentences from the Passage. You may add minimal connecting words (such as ``and'', ``which'', ``that'', ``because'', ``thus'') to make the answer grammatically correct, but do not paraphrase the original wording unless necessary. \\
\end{tcolorbox}
\caption{Answer Formatting Templates. Template 1 was presented in Ateia \& Kruschwitz's published work \cite{ateia2024can} and thus omitted. }
\label{fig:formatting-templates}
\end{figure}

We experimented with two primary dimensions in our prompting strategy: 
\begin{enumerate}
    \item \textbf{Number of Few-Shot Examples:} We tested using either a single example (1-shot) or ten examples (10-shots) as contextual demonstrations, corresponding to $n=1$ or $10$ in Fig. \ref{fig:prompt-structure}. The examples were drawn from the training set as snippets-question-answer triplets to guide the model in generating appropriate responses. 
    \item \textbf{Prompts Templates Design:} Each prompt varied along two core components: the \textbf{Query Template}, which defined how few-shot examples and questions were presented, and the \textbf{Answer Formatting Template}, which instructed the model on the expected format of question-specific responses. Below we describe the distinct styles we implemented:
    \begin{itemize}
        \item \textbf{Query Templates:} As shown in Fig. \ref{fig:query-templates}, Template 1 generates prompts for all exact answers and ideal answers independent of the question id. While in Template 2, we integrate the system-generated exact answers as a hint for non-summary-type ideal answers. Additionally, we modified the list type example: in Template 1, only a single correct list item for each entry was shown as example, while in Template 2, we show the entire golden answer.  
        \item \textbf{Answer Formatting Templates}: Template 1 was adapted from the instruction style proposed by Ateia \& Kruschwitz \cite{ateia2024can}, specifying concise, type-aware answers consistent with BioASQ guidelines. In Fig. \ref{fig:formatting-templates}, we illustrate the key modifications introduced in Template 2, which were designed to address observed issues in system outputs—such as redundancy, synonym repetition, or incomplete phrasing—especially for list and factoid types.
\end{itemize}

Combining different Query and Answer Formatting Templates, we defined three distinct prompt styles used in our experiments:
    \begin{itemize}
        \item \textbf{Style 1}: Query Template 1 + Answer Formatting Template 1. 
        \item \textbf{Style 2}: Query Template 2 + Answer Formatting Template 1. 
        \item \textbf{Style 3}: Query Template 2 + Answer Formatting Template 2.
    \end{itemize}
\end{enumerate}

\section{Results}


\subsection{Phase A: Retrieval}
The retrieval comparison method evaluates the performance of a vector embedding database against PubMed's retrieval system using BioASQ dataset, which contains 5,389 question - PMID gold standard pairs. We tested retrieving top 10, 100, 1000, and 10,000 documents from the vector database and measuring the counts of true positives as shown in Table \ref{tb:vector-db-vs-pubmed}

\begin{table}[t]
\centering
\caption{Retrieval Performance Comparison of PubMed Search API vs  our Vector Database. Recall@N computes the fraction of golden documents contained in the retrieved N documents.}
\label{tb:vector-db-vs-pubmed}
\begin{tabular}{lcccc}
\toprule
\textbf{Retrieval System} & \textbf{Recall@10} & \textbf{Recall@100} & \textbf{Recall@1k} & \textbf{Recall@10k} \\
\midrule
PubMed Search API & 0.10 & 0.13 & 0.23 & 0.25 \\
Vector DB w/ bge-large-en & 0.23 & 0.38 & 0.56 & 0.91 \\
\bottomrule
\end{tabular}
\end{table}

\subsection{Phase A: Re-Ranker}


We evaluated the re-rankers using the MAP@10 metric which is a popular measure for IR tasks balancing recall@10 as well as the ranking in top10, in addition to being the official leaderboard ranking metric for PhaseA of the competition. It is defined as:

\begin{displaymath}
\text{AP} = \frac{ \sum_{r=1}^{|L|} P(r) \cdot \text{rel}(r)}{min(|L_R|, 10)}
\end{displaymath}
\begin{displaymath}
\text{MAP} = \frac{1}{n} \sum_{i=1}^{n} \text{AP}_i
\end{displaymath}
where $|L|$ is the number of items in the list, $|L_R|$ is the number of relevant items, $P(r)$ is the precision when the returned list is treated as containing only its first $r$ items, and $rel(r)$ equals 1 if the r-th item of the list is in the golden set (i.e., if the r-th item is relevant) and 0 otherwise.

\subsubsection{Cross-Encoder Results}

\begin{table}[t]
\centering
\caption{MAP@10 scores of different models across index retrieval sizes. MAP scores are based on the last 100 questions in the dataset. Finetuning was done using our index retrieved documents and golden documents from provided dataset for 1 epoch. Last row is using the best MAP score model ms-marco-MiniLM-L12 finetuned for 3 epochs.}
\label{tab:reranker_map_scores}
\begin{tabular}{lccccc}
\toprule
 & \multicolumn{5}{c}{\textbf{Num Docs Retrieved from Index}} \\
\cmidrule(lr){2-6}
\textbf{Re-Ranker Model} & \textbf{Top 10} & \textbf{Top 100} & \textbf{Top 200} & \textbf{Top 500} & \textbf{Top 1k} \\
\midrule
Retrieve top10 \& No Re-ranking & 0.1895 & - & - & - & - \\
\midrule
ms-marco-MiniLM-L-6 & 0.2347 & 0.3048 & 0.3081 & 0.3130 & 0.3179 \\
(\textit{Finetuned}) ms-marco-MiniLM-L-6 & 0.2454 & 0.3543 & 0.3671 & 0.3864 & 0.3917 \\
ms-marco-MiniLM-L12 & 0.2305 & 0.3062 & 0.3023 & 0.3075 & 0.3130 \\
(\textit{Finetuned}) ms-marco-MiniLM-L12 & 0.2521 & 0.3684 & 0.3736 & 0.3842 & \textbf{0.3941} \\
jina-reranker-v2-base-multilingual & 0.2424 & 0.3202 & 0.3266 & 0.3445 & 0.3533 \\
(\textit{Finetuned}) jina-reranker-v2-base-multilingual & 0.2430 & 0.3437 & 0.3519 & 0.3707 & 0.3785 \\
(\textit{Finetuned}) bge-reranker-large & 0.2250 & 0.3124 & 0.3186 & 0.3382 & 0.3388 \\
\midrule
(\textbf{\textit{Finetuned 3epochs}) ms-marco-MiniLM-L12} & 0.2523 & 0.3972 & 0.4064 & 0.4193 & \textbf{0.4337} \\
\bottomrule
\end{tabular}
\end{table}

We tried a few cross-encoders in their vanilla form as well as after finetuning them on our dataset. We kept upto 1k retrieved documents with $\approx$10 golden documents for each question in our dataset which was split into train, val and test sets stratified by question. The model performance was assessed based on the ranking metric MAP@10 on validation split. Once we had the finetuned models, we did a final test of their performance (again using MAP@10) for different number of retrievals from the index as shown in Table \ref{tab:reranker_map_scores}. This allowed us to select the optimal retrieval count from index (to tradeoff better recall with high retrieval vs better ranking with low retrieval).

A general trend we observed is that each model performs better after finetuning with an increase of 0.02 - 0.08 in the MAP@10 score. Each of these models were finetuned for only 1 epoch to establish the best performing model which was "\textit{1k document retrieval followed by finetuned ms-marco-MiniLM-L12 for re-ranking}" with MAP@10 of 0.3941. This best model was then finetuned for longer (3 epochs) leading to MAP@10 of \textbf{0.4337}. This is the cross-encoder that we used in all our systems for our submission.

\subsubsection{GPT and Ensembling Results}
With the nomination approach, where first 6 documents came from the finetuned ms-marco-MiniLM-L12 cross encoder and the remaining 4 spots were filled by the GPT-4o re-ranker, we were able to increase the MAP@10 by only 0.0073.

With weighted ensemble of the 2 re-rankers a) finetuned ms-marco-MiniLM-L12, b) finetuned ms-marco-MiniLM-L12 to get top30 followed by gpt-based re-ranker to get top10, we were able to achieve our best MAP@10 of \textbf{0.4551} with weights of 1 and 7 respectively for the two re-rankers. As mentioned previously, we arrived at those weights by running grid search on the validation dataset.

\subsection{Phase A: Leaderboard}
For PhaseA, we participated only in batch 4, the results of which are shown in Table \ref{tab:phasea-leaderboard}. Our best performing system achieved Rank 10 with MAP@10 of 0.1581. As expected, the ensemble model performed the best followed by system4 which used a two-stage re-ranker (finetuned cross encoder to get top30 from 1k, then gpt-based re-ranker to get to final top10).
The remaining 3 systems all used different retrieval counts and re-ranking with a single finetuned cross encoder. As expected retrieving 2k performs better than retrieving 1k, but retrieving 5k actually performs worse than the rest. This suggests that there's a turning point somewhere b/w 2k and 5k where the increase in recall vs more docs to re-rank tradeoff flips.
\begin{table}[t]
\centering
\caption{Leaderboard for batch 4 of Phase A showing MAP@10 scores and non-dense ranks of top performer and our submissions. Our best performing system retrieved 1k documents from index and re-ranked them using a weighted ensemble of finetuned cross encoder ms-marco-MiniLM-L12 and gpt-based re-ranker. This submission achieved a rank of 10 (and dense rank of 9) out of 78 total submissions in the document retrieval task.}
\label{tab:phasea-leaderboard}
\begin{adjustbox}{width=\textwidth}
\begin{tabular}{c l l c}
\toprule
\textbf{Rank} & \textbf{System} & \textbf{Description} & \textbf{MAP@10} \\
\midrule
1   & bioinfo-1            & Top submission                   & 0.1801 \\
...   & ...            & ...                   & ... \\
10  & \textbf{DS@GTBioASQT13b3}     & Retrieve 1k, Ensemble FT Cross Encoder \& GPT Reranker                     & \textbf{0.1581} \\
11  & \textbf{DS@GTBioASQT13b4}     & Retrieve 1k, FT Cross Encoder (top30) -> GPT Reranker (top10)                     & \textbf{0.1562} \\
12  & Using KG for list q  & -           & 0.1476 \\
13  & Main pipeline        & -           & 0.1466 \\
14  & dmiip2024           & -                     & 0.1433 \\
15  & dmiip2024\_1        & -               & 0.1423 \\
16  & \textbf{DS@GTBioASQT13b2}   & Retrieve 2k, FT Cross Encoder (top10)                 & \textbf{0.1337} \\
17  & IR4                 & -                    & 0.1333 \\
18  & \textbf{DS@GTBioASQT13b1}   & Retrieve 1k, FT Cross Encoder (top10)                 & \textbf{0.1331} \\
19  & \textbf{DS@GTBioASQT13b5}    & Retrieve 5k, FT Cross Encoder (top10)                     & \textbf{0.1309} \\
\bottomrule
\end{tabular}
\end{adjustbox}
\end{table}

\subsection{Phase B: Answer Generation}
While performance on previous batch submissions were not available during system development, we randomly sampled 80 questions from the training set—20 for each question type—and constructed queries using the gold-standard snippets. This simulation-based evaluation guided us to design system for better performance in real answer generation.


We briefly introduce the evaluation metrics used for each answer type below. Further details are available in the official BioASQ documentation \cite{BioASQ_metrics_guide}:
\begin{itemize}
    \item \textbf{Yes/No Questions}: Macro-averaged $F_1$ score (maF1):  $\text{maF1} = (F_{1y} + F_{1n})/2$.
    \item \textbf{Factoid Questions}: Mean Reciprocal Rank (MRR): Average inverse rank of the first correct entity.
    \item \textbf{List Questions}: Mean $F_1$ (mF1) score.
    \item \textbf{Summary Questions/Ideal Answers}: Mean ROUGE-2 (mROUGE-2) and ROUGE-SU4 metrics that measure overlap between two texts.
\end{itemize}

Table \ref{tb:PhaseB:eval:exact} and \ref{tb:PhaseB:eval:ideal} summarize model performance across exact and ideal answer types under different prompt styles. Additionally, we present batch submission results and corresponding performance rankings against other teams in Table \ref{tab:exact-ranks} (exact answers) and Table \ref{tab:ideal-rankes} (ideal answers). Note that we focused on prompt style comparison rather than number of examples after we observe no significant improvement 1-shot vs. 10-shots results with prompt style 1.

\begin{table}[t]
\centering
\caption{Performance comparison between GPT-4o-turbo and Mistral-7B-Instruct-v0.3 across different prompt styles and question types, using official evaluation metrics for exact answers.}
\label{tb:PhaseB:eval:exact}
\begin{adjustbox}{width=\textwidth}
\begin{tabular}{l c c c c c c}
\toprule
\textbf{Prompt Setting} & \multicolumn{3}{c}{\textbf{GPT-4o-turbo}} & \multicolumn{3}{c}{\textbf{Mistral-7B-Instruct-v0.3}} \\
\cmidrule(lr){2-4} \cmidrule(lr){5-7}
 & Yes/No (maF1) & Factoid (MRR) & List (mF1) & Yes/No (maF1) & Factoid (MRR) & List (mF1) \\
\midrule
1-shot + style 1   & 1.0 & 0.40 & 0.59 & 1.0 & 0.30 & 0.54 \\
10-shots + style 1 & 1.0 & 0.45 & 0.59 & 1.0 & 0.15 & 0.50 \\
1-shot + style 2   & 1.0 & 0.46 & 0.64 & 1.0 & 0.25 & 0.56 \\
1-shot + style 3   & 1.0 & 0.35 & 0.62 & 1.0 & 0.25 & 0.64 \\
\bottomrule
\end{tabular}
\end{adjustbox}

\centering
\caption{Performance comparison between GPT-4o-turbo and Mistral-7B-Instruct-v0.3 across different prompt styles on summary/ideal answers with mROUGE-2 metrics.}
\label{tb:PhaseB:eval:ideal}
\begin{adjustbox}{width=0.8\textwidth}
\begin{tabular}{l c c c c c c}
\toprule
\textbf{Prompt Setting} & \multicolumn{3}{c}{\textbf{GPT-4o-turbo}} & \multicolumn{3}{c}{\textbf{Mistral-7B-Instruct-v0.3}} \\
\cmidrule(lr){2-4} \cmidrule(lr){5-7}
 & Recall & Precision & F1-measure & Recall & Precision & F1-measure \\
\midrule
1-shot + style 1   & 0.22 & 0.19 & 0.19 & 0.30 & 0.20 & 0.22 \\
10-shots + style 1 & 0.23 & 0.19 & 0.19 & 0.31 & 0.24 & 0.26 \\
1-shot + style 2   & 0.20 & 0.22 & 0.19 & 0.28 & 0.19 & 0.21 \\
1-shot + style 3   & 0.30 & 0.32 & 0.28 & 0.28 & 0.19 & 0.21 \\
\bottomrule
\end{tabular}
\end{adjustbox}

\end{table}

Focusing on exact answers (Table \ref{tb:PhaseB:eval:exact}), both GPT-4o-turbo and Mistral-7B-Instruct-v0.3 achieved perfect or near-perfect maF1 scores on yes/no questions across all settings, indicating reliable performance for binary decision with prompting. However, the real submission performance dropped in batch 4, suggesting room for improvement. 

For factoid questions, GPT-4o-turbo consistently outperformed Mistral, with MRR scores ranging from 0.35 to 0.46, compared to 0.15 to 0.30 for Mistral. Interestingly, the systems performs much better on real submissions, with a round 0.2 increase in MRR for both LLM models. The GPT models with Style 2 and 3 (described in previous sections) ranked 1st and 2nd on the batch 4 submission. In batch 4, GPT models using styles 2 and 3 ranked 1st and 2nd, respectively. Interestingly, GPT4-1shot-Style1 achieved similar MRRs in both batch 2 and batch 4, yet its ranking improved significantly from 23rd to 8th, suggesting that our system has greater robustness across question sets, staying consistent in performance. However, the prompt style change do not show significant impact on factoid questions. This trend held true in both simulated and real submission results.

In contrast, prompt styles had a more pronounced effect on list questions. In our experiment, both models showed gains in mean F1 (mF1) when using styles 2 and 3 over style 1. GPT-4o-turbo achieved an mF1 of 0.64, with Mistral also showing modest improvements. These gains were echoed in batch results: the Mistral7BIns10shots system improved its list F1 and rank from 64th/62nd in batches 1/2 (style 1) to 33rd in batch 4 (style 3). Similarly, GPTPrompt1sStyle3 ranked 5th in batch 4, compared to 14th/15th with style 1/2, confirming the importance of prompt design in enhancing list-question performance (Table \ref{tab:exact-ranks}).

For ideal answers, both models achieved modest mROUGE-2 scores across all prompt configurations (Table \ref{tb:PhaseB:eval:ideal}) . Transitioning from style 1 to style 2—which integrated the exact answer into the prompt—did not lead to meaningful gains for either model. However, for GPT-4o-turbo, style 3 yielded measurable improvements in summary F1 scores. This trend is also reflected in batch 4 results (Table \ref{tab:ideal-rankes}), where GPT-4o-turbo with style 3 surpassed Mistral in ROUGE-SU4 ranking.

Interestingly, the integration of the exact answer in style 2 did not improve summarization for either model, suggesting that the main limitation was not content selection, but rather expression. The improvement observed with style 3 highlights a limitation of ROUGE metrics: they focus on surface-level n-gram overlap rather than semantic equivalence. Style 3 explicitly instructed the model to``only combining direct phrases or sentences from the Passage'' and to “avoid paraphrasing the original wording unless necessary,” which increased overlap with the gold snippets and thus boosted ROUGE scores. This also explained why Mistral constantly outperformed GPT-4o-turbo in ROUGE-based evaluations. The larger GPT model may generate more flexibly paraphrased responses from the context compared to Mistral-7B-Instruct-v0.3, thus penalizing the ROUGE scores despite producing semantically correct answers.

In summary, GPT-4o-turbo showed stronger performance overall on exact answers, especially for factoid and list questions, while Mistral-7B-Instruct-v0.3 was more competitive in summarization. Prompt style mattered more for list and ideal answers, and carefully chosen prompting strategies can help unlock the strengths of each model.

\begin{table}[t]
\centering
\caption{Leaderboard results for Phase B showing evaluation metric scores of various systems on exact answers: mF1 (Yes/No), MRR (Factoid), and F1-measure (List), with corresponding non-dense ranks (1 = best, underscored rank means tie). In batch 4, our submission achieved a dense rank of 3 for yes/no questions, dense rank of 1 for factoid questions and dense rank of 5 for list questions with a total of 79 submissions on the leaderboard.}
\label{tab:exact-ranks}
\begin{adjustbox}{width=\textwidth}
\begin{tabular}{c l l cc cc cc }
\hline
  &   &  & 
\multicolumn{2}{c}{\textbf{Yes/No}} & 
\multicolumn{2}{c}{\textbf{Factoid}} & 
\multicolumn{2}{c}{\textbf{List}} \\
\cmidrule(lr){4-5} \cmidrule(lr){6-7} \cmidrule(lr){8-9}
\textbf{Batch} & \textbf{Presented Name} & \textbf{Actual Style} & 
\textbf{maF1} & \textbf{Rank} & 
\textbf{MRR} & \textbf{Rank} & 
\textbf{F1-measure} & \textbf{Rank} \\
\hline
1 & Mistral7BIns10shots & Mistral-10shots-Style1 & 1.00 & \underline{\textbf{1}} & 0.37 & 55 & 0.27 & 64 \\
\midrule
2 & Mistral7BIns10shots & Mistral-10shots-Style1 & 1.00 & \underline{\textbf{1}} & 0.15 & 61 & 0.23 & 62 \\
2 & GPT4turbo            & GPT4-1shot-Style1     & 1.00 & \underline{\textbf{1}} & 0.56 & 23 & 0.52 & 24 \\
\midrule
4 & GPT4turbo            & GPT4-1shot-Style1 & 0.95 & \underline{12} & 0.57 & \underline{\textbf{8}} & 0.58 & 14 \\
4 & GPTPrompt1sStyle2    & GPT4-1shot-Style2 & 0.95 & \underline{12} & 0.64 & \textbf{1} & 0.58 & 15 \\
4 & GPTPrompt1sStyle3    & GPT4-1shot-Style3 & 0.95 & \underline{12} & 0.61 & \textbf{2} & 0.63 & \textbf{5} \\
4 & Mistral7BIns10shots  & Mistral-10shots-Style3 & 0.82 & 56 & 0.48 & \underline{38} & 0.51 & 33 \\
\hline
\end{tabular}
\end{adjustbox}
\end{table}

\begin{table}[t]
\centering
\caption{Leaderboard results for Phase B on ideal answers based on ROUGE metrics: ROUGE-2 (Recall/F1) and ROUGE-SU4 (Recall/F1), with corresponding non-dense ranks (1 = best). In batch 4, our submission achieved a dense and non-dense rank of 12 based on ROUGE-2 (F1) metric and a dense as well as non-dense rank of 11 based on ROUGE-SU4 (F1) metric with a total of 73 submissions on the leaderboard.}
\label{tab:ideal-rankes}
\begin{adjustbox}{width=0.9\textwidth}
\begin{tabular}{c l c 
                c c 
                c c 
                c c }
\toprule
  &  &  & 
\multicolumn{3}{c}{\textbf{ROUGE-2}}  & 
\multicolumn{3}{c}{\textbf{ROUGE-SU4}} \\
\cmidrule(lr){4-6} \cmidrule(lr){7-9} 
\textbf{Batch} & \textbf{System} & \textbf{Style} & 
\textbf{Recall} & \textbf{F1} & \textbf{Rank} & 
\textbf{Recall} & \textbf{F1} & \textbf{Rank} \\
\midrule
1 & Mistral7BIns10shots & Mistral-10shots-Style1 & 0.35 & 0.26 & 16 & 0.34 & 0.25 & 16 \\
\midrule
2 & Mistral7BIns10shots & Mistral-10shots-Style1 & 0.36 & 0.32 & 12 & 0.35 & 0.30 & 12 \\
2 & GPT4turbo            & GPT4-1shot-Style1      & 0.31 & 0.28 & 16 & 0.31 & 0.27 & 15 \\
\midrule
4 & GPT4turbo            & GPT4-1shot-Style1      & 0.27 & 0.23 & 22 & 0.26 & 0.23 & 21 \\
4 & GPTPrompt1sStyle2    & GPT4-1shot-Style2      & 0.24 & 0.23 & 26 & 0.24 & 0.23 & 22 \\
4 & GPTPrompt1sStyle3    & GPT4-1shot-Style3      & 0.28 & 0.29 & 23 & 0.28 & 0.29 & 11 \\
4 & Mistral7BIns10shots  & Mistral-10shots-Style3 & 0.36 & 0.29 & 12 & 0.35 & 0.28 & 14 \\
\bottomrule
\end{tabular}
\end{adjustbox}
\end{table}

\section{Discussion}


Our custom Qdrant Vector Database, utilizing the bge-large-en bi-encoder, significantly outperformed the PubMed Query API in retrieving relevant documents. As shown in Table \ref{tb:vector-db-vs-pubmed}, all Recall@N values (e.g., Recall@10 = 0.23 vs. 0.10, Recall@1k = 0.56 vs. 0.23) demonstrate the superiority of dense-vector search over traditional keyword-based retrieval. This advantage stems from the bi-encoder’s ability to capture semantic relationships in biomedical texts.

Initial retrieval of 1k candidates followed by re-ranking with an ensemble of cross-encoders and GPT-based re-rankers, further enhanced the retrieval performance as shown in Table \ref{tab:reranker_map_scores}. While retrieving top10 documents from index with no re-ranker gets us an MAP@10 of 0.1895 only, retrieving 1k followed by re-ranking via finetuned cross encoder achieves an MAP@10 of 0.4337 (more than doubled). Applying ensemble re-rankers takes us all the way to \textbf{0.4551} on our test set. This reflects the effectiveness of combining bi-encoders for high recall with cross-encoders and LLMs for precise re-ranking, allowing the system to identify the most relevant documents from a large pool. It is interesting to note that each next stage is as much or more accurate as well as more expensive on the IR task. Bi-encoders are the most efficient, seamlessly searching through millions of documents to get us an initial 1k candidates. Cross-encoders jointly look at query and candidate pair making them slow albeit more accurate, which is why they are applied on only the smaller pool of retrieved 1k documents. Using LLMs as re-rankers takes more time to re-rank and is much more costly per question but probably slightly more accurate than cross-encoders, which is why they are employed on the top30 cross-encoder re-ranked documents only.

An interesting observation in the re-ranking stage (Table \ref{tab:reranker_map_scores}) is that models like jina-reranker-v2-base-multilingual and bge-reranker-large are much larger models with 278M and 560M params respectively compared to the best performing model ms-marco-MiniLM-L12 which has only 33.4M params. This reveals the fact that the pre-training of each model and the model's architecture lead to models specializing in different aspects and tasks.

As a future followup, we would like to explore using language models to reformulate the query for improved retrievals. Additionally, increasing the retrieval from 1k to 2k showed improvement (System2 had better score than System1), which means we can further optimize the configuration for our best system by retrieving more documents from the index. 

For Phase B, we also experimented with model fine-tuning using BioGPT. As a decoder-only model trained to generate continuous text based on prior context, BioGPT inherently outputs free-form text. For instance, when responding to a yes/no question, it frequently provided verbose answers instead of direct “yes” or “no” responses. To rectify this, we implemented task-specific heads on top of the pre-trained model: a linear layer projecting hidden states into binary outputs for yes/no questions and a span-prediction head for identifying the start and end indices of answers within provided snippets for factoid questions. Due to limited training data and computational constraints, we anticipated limited performance improvements. Additionally, memory constraints and unstable gradient descent posed challenges. Despite resolving stability issues, fine-tuned methods consistently showed limited performance: yes/no predictions defaulted to “yes,” and factoid responses frequently were incomplete. Consequently, results presented herein focus solely on prompting strategies, though detailed fine-tuning outcomes are publicly available in our GitHub repository.

Returning to our submitted prompting-based Phase B systems, although these systems achieved perfect scores for yes/no questions in our internal experiments, the real submission batches continued to hold the success in batch1 and batch2 but dropped a bit for batch4. This discrepancy indicates a critical limitation of our evaluation strategy: sampled evaluation questions may not comprehensively represent the complexity and variability encountered in real-world questions, masking shortcomings in our prompts for yes/no questions. Moreover, certain questions consistently underperformed across all tested systems, often due to overly long or repetitive snippets that diluted contextual relevance. These observations suggest future research directions such as introducing snippet filtering or prioritization methods to select more diverse and informative contexts, potentially enhancing both exact and ideal answer generation.

\section{Conclusion}
In this paper, we discussed our participation in the BioASQ's Semantic Question Answering challenge. We developed a RAG system for biomedical queries which retrieves from a huge PubMed index (O(10M) documents). We have shown that our custom local Qdrant Vector Database with Bi-Encoder performs significantly better than using PubMed Query API. The retrieval performance was further improved by employing a two-stage re-ranker ensemble. We got competitive results on the leaderboard at rank 10 and we believe that this tiered approach is the way to go for future improvements.

In answer generation phase, we were able to top the leaderboard for factoid questions and be competitive for all other question types (generally in top10), showing that for generating real natural language responses, instruction tuned LLMs are extremely competitive.

A Biomedical Semantic Question Answering system has huge implications in the real-world. This can help medical researchers in the field as well as layman users in answering their queries with authoritative sources to back them up. It can also help mitigate medical misinformation by applying to fact-check various claims on the internet.

\section*{Acknowledgements}

We thank the Data Science at Georgia Tech (DS@GT) CLEF competition group for their support.

\section*{Declaration on Generative AI}
 During the preparation of this work, the author(s) used ChatGPT in order to: properly format tables and reword certain lines. After using this tool/service, the author(s) reviewed and edited the content as needed and take(s) full responsibility for the publication’s content.

\bibliography{main}




\end{document}